\documentclass[useAMS,usegraphicx,usenatbib]{mn2e}

\usepackage{times}
\usepackage{amssymb}
\usepackage{amsmath}

\begin{document}

\include{defn}
\def\cm{{\rm\thinspace cm}}
\def\gm{{\rm\thinspace gm}}
\def\dyn{{\rm\thinspace dyn}}
\def\erg{{\rm\thinspace erg}}
\def\eV{{\rm\thinspace eV}}
\def\MeV{{\rm\thinspace MeV}}
\def\g{{\rm\thinspace g}}
\def\ga{{\rm\thinspace gauss}}
\def\K{{\rm\thinspace K}}
\def\keV{{\rm\thinspace keV}}
\def\km{{\rm\thinspace km}}
\def\kpc{{\rm\thinspace kpc}}
\def\Lsun{\hbox{$\rm\thinspace L_{\odot}$}}
\def\m{{\rm\thinspace m}}
\def\Mpc{{\rm\thinspace Mpc}}
\def\Msun{\hbox{$\rm\thinspace M_{\odot}$}}
\def\Zsun{\hbox{$\rm\thinspace Z_{\odot}$}}
\def\pc{{\rm\thinspace pc}}
\def\ph{{\rm\thinspace ph}}
\def\s{{\rm\thinspace s}}
\def\yr{{\rm\thinspace yr}}
\def\sr{{\rm\thinspace sr}}
\def\Hz{{\rm\thinspace Hz}}
\def\MHz{{\rm\thinspace MHz}}
\def\GHz{{\rm\thinspace GHz}}
\def\chisq{\hbox{$\chi^2$}}
\def\delchi{\hbox{$\Delta\chi$}}
\def\cmps{\hbox{$\cm\s^{-1}\,$}}
\def\cmpssq{\hbox{$\cm\s^{-2}\,$}}
\def\cmsq{\hbox{$\cm^2\,$}}
\def\cmcu{\hbox{$\cm^3\,$}}
\def\pcmcu{\hbox{$\cm^{-3}\,$}}
\def\pcmcuK{\hbox{$\cm^{-3}\K\,$}}
\def\dynpcmsq{\hbox{$\dyn\cm^{-2}\,$}}
\def\ergcmcups{\hbox{$\erg\cm^3\ps\,$}}
\def\ergpcmps{\hbox{$\erg\cm^{-3}\s^{-1}\,$}}
\def\ergpcmsqps{\hbox{$\erg\cm^{-2}\s^{-1}\,$}}
\def\ergpcmsqpspA{\hbox{$\erg\cm^{-2}\s^{-1}$\AA$^{-1}\,$}}
\def\ergpcmsqpspsr{\hbox{$\erg\cm^{-2}\s^{-1}\sr^{-1}\,$}}
\def\ergpcmcups{\hbox{$\erg\cm^{-3}\s^{-1}\,$}}
\def\ergpcmps{\hbox{$\erg\cm^{-1}\s^{-1}$}}
\def\ergps{\hbox{$\erg\s^{-1}\,$}}
\def\ergpspmp{\hbox{$\erg\s^{-1}\Mpc^{-3}\,$}}
\def\gpcm{\hbox{$\g\cm^{-3}\,$}}
\def\gpcmps{\hbox{$\g\cm^{-3}\s^{-1}\,$}}
\def\gps{\hbox{$\g\s^{-1}\,$}}
\def\Jy{{\rm Jy}}
\def\keVpcmsqpspsr{\hbox{$\keV\cm^{-2}\s^{-1}\sr^{-1}\,$}}
\def\kmps{\hbox{$\km\s^{-1}\,$}}
\def\kmpspmp{\hbox{$\km\s^{-1}\Mpc{-1}\,$}}
\def\Lsunppc{\hbox{$\Lsun\pc^{-3}\,$}}
\def\Msunpc{\hbox{$\Msun\pc^{-3}\,$}}
\def\Msunpkpc{\hbox{$\Msun\kpc^{-1}\,$}}
\def\Msunppc{\hbox{$\Msun\pc^{-3}\,$}}
\def\Msunppcpyr{\hbox{$\Msun\pc^{-3}\yr^{-1}\,$}}
\def\Msunpyr{\hbox{$\Msun\yr^{-1}\,$}}
\def\pcm{\hbox{$\cm^{-3}\,$}}
\def\pcmsq{\hbox{$\cm^{-2}\,$}}
\def\pcmK{\hbox{$\cm^{-3}\K$}}
\def\phpcmsqps{\hbox{$\ph\cm^{-2}\s^{-1}\,$}}
\def\pHz{\hbox{$\Hz^{-1}\,$}}
\def\pmpc{\hbox{$\Mpc^{-1}\,$}}
\def\pmpccu{\hbox{$\Mpc^{-3}\,$}}
\def\ps{\hbox{$\s^{-1}\,$}}
\def\psqcm{\hbox{$\cm^{-2}\,$}}
\def\psr{\hbox{$\sr^{-1}\,$}}
\def\kmpspMpc{\hbox{$\kmps\Mpc^{-1}$}}
\def\apj{ApJ}
\def\mnras{MNRAS}
\def\nat{Nat}
\def\physrevB{Phys. Rev. B}
\def\araa{ARA\&A}                
\def\aap{A\&A}                   
\def\aaps{A\&AS}                 
\def\aj{AJ}                      
\def\apjs{ApJS}                  
\def\pasp{PASP}                  
\def\apjl{ApJ}                   
\def\pasj{PASJ}

\voffset=-0.4in

\title[Radiation pressure and absorption in AGN]{The effect of
radiation pressure on dusty absorbing gas around AGN}
\author[A.C. Fabian, R.V. Vasudevan \& P. Gandhi]
{\parbox[]{6.in} {A.C. Fabian$^1$, R.V. Vasudevan$^1$ and P. Gandhi$^2$\\
\footnotesize
$^1$Institute of Astronomy, Madingley Road, Cambridge CB3 0HA\\
$^{2}$RIKEN Cosmic Radiation Lab, 2-1 Hirosawa, Wakoshi, Saitama 351-0198, Japan\\}}

\maketitle 

\begin{abstract} 
  Many Active Galactic Nuclei (AGN) are surrounded by gas which
  absorbs the radiation produced by accretion onto the central black
  hole and obscures the nucleus from direct view. The dust component
  of the gas greatly enhances the effect of radiation pressure above
  that for Thomson scattering so that an AGN which is sub-Eddington
  for ionized gas in the usual sense can appear super-Eddington for
  cold dusty gas.  The radiation-pressure enhancement factor depends
  on the AGN spectrum but ranges between unity and about 500,
  depending on the column density. It means that an AGN for which the
  absorption is long-lived should have a column density $N_{\rm H}>
  5\times 10^{23}\lambda \psqcm$, where $\lambda$ is its Eddington
  fraction $L_{\rm bol}/L_{\rm Edd}$, provided that $N_{\rm H}>5\times
  10^{21}\psqcm$. We have compared the distribution of several samples
  of AGN -- local, CDFS and Lockman Hole -- with this expectation and
  find good agreement. We show that the limiting enhancement factor
  can explain the black hole mass -- bulge mass relation and note that
  the effect of radiation pressure on dusty gas may be a key component
  in the feedback of momentum and energy from a central black hole to
  a galaxy.

\end{abstract} 

\section{Introduction}

Active Galactic Nuclei are powered by accretion onto a central massive
black hole. The gas surrounding the nucleus, some of which provides
the fuel for the accretion process, often obscures it from direct
view. Indeed, the hard shape of the spectrum of the X-ray Background
argues that most accretion onto galactic nuclei is obscured
\protect\citep{1999MNRAS.303L..34F}. This is confirmed by deep Chandra
and XMM-Newton imaging of the Sky with many AGN found to lie behind a
column density of $10^{22} - 10^{23} \pcmsq$ or more
(\protect\citealt{2002ApJS..139..369G}; \protect\citealt{2005ARA&A..43..827B}).

Much of the obscuring material must lie within the inner 100~pc, or
its total mass would be prohibitive (see
\protect\citealt{2007astro.ph..1109M} for a review). It is therefore
part of the inner bulge of the host galaxy. Stars can also form from this
gas, giving a nuclear starburst. The gas is subject to the radiation
pressure of the AGN, and can be ejected from the bulge if the nucleus
becomes too bright. Such AGN feedback may thereby remove the gas which
fuels the nucleus and from which new stars form, so terminating the
growth of both the central black hole and its host bulge. Simple
calculations of when this occurs
(\protect\citealt{1999MNRAS.308L..39F};
\protect\citealt{2002MNRAS.329L..18F};
\protect\citealt{2003ApJ...596L..27K},
\protect\citealt{2005ApJ...618..569M};
\protect\citealt{2006MNRAS.373L..16F}) lead to the following relation
between the mass of the black hole $M_{\rm BH}$ and the velocity
dispersion of the bulge $\sigma$:
\begin{equation} 
M_{\rm BH}={{f\sigma^4 }\over{\pi G^2
      m_{\rm p}}}\sigma_{\rm T}.
\end{equation}
Assuming a gas fraction $f\sim 0.1$, this gives an $M_{\rm BH} -
\sigma$ relation in good agreement with observations
(\protect\citealt{1995ARA&A..33..581K};
\protect\citealt{1998AJ....115.2285M};
\protect\citealt{2000ApJ...539L..13G};
\protect\citealt{2001ApJ...555L..79F}).

The limit when radiation pressure ejects mass is an effective
Eddington limit relying on absorption of radiation by dust, not
electron scattering. The radiation pressure is amplified or boosted by
a factor $A$ which is the ratio of the effective, frequency weighted,
absorption cross section for dusty gas, $\sigma_{\rm d}$ to that for
electrons alone: $\sigma_{\rm T}$. 
\begin{equation}
A={\sigma_{\rm d}\over\sigma_{\rm T}}.
\end{equation}
The dust absorption is greatest in the UV and the value of $A$ depends
on the spectrum of the AGN. The X-ray emission from the nucleus
ensures that the gas and dust remain weakly ionized, and are
effectively coupled by Coulomb forces so that pressure on the grains
is shared with the surrounding gas .

Boost factors computed for a standard AGN spectrum, using the
radiation code {\sc cloudy} are shown in
\protect\cite{2006MNRAS.373L..16F} and Fig.~1. They
range from several hundred for low column densities and drop as the
column density of gas increases until they approach unity when the gas
becomes Compton thick, i.e. $N_{\rm H}\sim 1/{\sigma_{\rm T}} =
1.5\times 10^{24}\psqcm$. Under the assumptions used, the main
reduction of $A$ with column density $N_{\rm H}$ is merely due to the
increased column acting as a dead weight since the UV part of the
spectrum is used up after traversing a column of a few
$10^{21}\pcmsq$. The outer 'dead weight' is then pushed by the inner
gas which experiences the full force of the radiation. 

\begin{figure}
\includegraphics[width=\columnwidth]{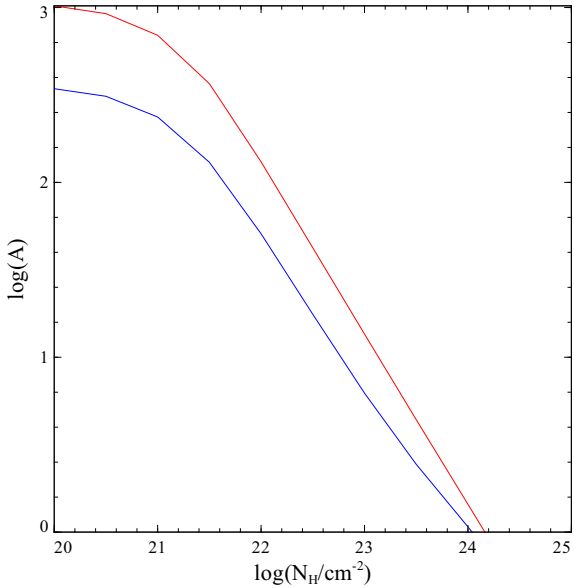}
\caption{The radiation pressure boost factor $A$ for dusty gas shown
  as a function of column density $N_{\rm H}$. The continuum spectrum
  is assumed to be that of high and low Eddington ratio objects (from
  \protect\cite{2007MNRAS.381.1235V}) for
  the upper and lower lines, respectively. }
\end{figure}

In this paper we explore further the implications of the boost factor on
the limit to the column density of dusty gas that can remain close to an
AGN without being ejected by radiation pressure. The key feature is to
cast the inverse of the boost factor as the (classical) Eddington ratio $\lambda$:
\begin{equation}
A^{-1} = {L_{\rm Bol}\over {L_{\rm Edd}}}=\lambda.
\end{equation}
A cloud is long lived if $A\lambda<1$. If the boost factor for a
particular cloud is 100 then it will be blown away when the Eddington
ratio exceeds 1/100. This determines a relation between the column
density of long-lived absorbing gas close to a nucleus and its
Eddington ratio.

\cite{1992ApJ...399L..23P} have previously noted the role of the Eddington
ratio when explaining the geometrical thickness of the torus around
AGN by the action of radiation pressure. \cite{2007MNRAS.380.1172H}
have estimated the effective Eddington limit for a torus around an AGN
considering both smooth and clumpy dust distributions.  

Previous work on the luminosity dependence of absorption has found
that the incidence of absorption drops with luminosity
(\protect\citealt{2003ApJ...598..886U};
\protect\citealt{2005ApJ...635..864L}, but see
\protect\citealt{2006MNRAS.372.1755D},
\protect\citealt{2005ApJ...630..115T}). The 'receding torus' model
(\protect\citealt{1991MNRAS.252..586L};
\protect\citealt{1998MNRAS.297L..39S}) is often  invoked to explain why
quasars may show less absorption. In that model the reduction in
absorbed objects is attributed to (heating) destruction of dust within
the inner regions of a torus of absoring gas surrounding the nucleus.

Here we consider the relation between luminosity and absorption from
the point of view of the effective Eddington limit on dusty gas.

\section{The relation between Boost Factor and Eddington Ratio}

We determine the boost factor $A$ as detailed in
\protect\cite{2006MNRAS.373L..16F} using {\sc cloudy} and AGN spectral
energy distributions obtained by
\protect\cite{2007MNRAS.381.1235V}. $A$ is obtained from absorption
only (the input spectrum minus the transmitted one) and assumes that
the gas is optically thin to the infrared radiation thereby produced.
Trapping of radiation is assumed to be negligible and the ionization
parameter $\xi=L/nr^2$ is arranged to be about 10 (thereby fixing the
gas density for a given incident AGN
flux). \protect\cite{2007MNRAS.381.1235V} find that the UV -- X-ray
spectral energy distributions (SEDs) of AGN depend on Eddington ratio,
with much more ionizing radiation -- and therefore higher boost
factors -- occurring at higher $\lambda$. We adopt mean SEDs for high
($>0.1$) and low ($<0.1$) $\lambda$ when computing $A$ as a function
of absorption column density. $N_{\rm H}$ is plotted against
$\lambda=A^{-1}$ in Fig.~2. The drift velocity of the grains relative
to the gas is computed by {\sc cloudy} and found to be low (about
$0.1\kmps$ where most of the radiative force is applied) justifying
our assumption that the dust and gas are coupled.

\begin{figure}
\includegraphics[width=\columnwidth]{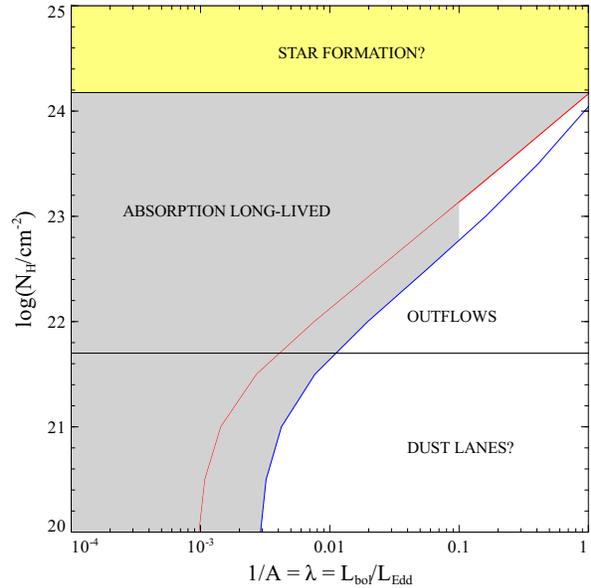}
\caption{Fig.~1 replotted in terms of column density versus
  $1/A$. Long-lived absorption from clouds near the centre of the
  galaxy should occur in the shaded region. Absorption from clouds and
  dust lanes can occur at higher values of $\lambda$ progressively
  further out. High column densities there would require prohibitive
  gas masses so we restrict such a region to $N_{\rm
    H}<5\times 10^{21}\psqcm,$ marked by a horizontal line. }
\end{figure}

Long-lived absorbing clouds can survive against radiation pressure in
the shaded region of the Figure. Clouds to the right of the dividing
line, in the unshaded part, see the nucleus as above the effective
Eddington luminosity and are thus ejected. Objects found in this
region should be experiencing outflows and absorption may be transient
or variable. The Compton-thick objects never see the source as
exceeding the Eddington luminosity and so can be long-lived at all
Eddington ratios less than unity. Gas clouds in this regime are more
likely to change because of star formation.

Absorbed objects in the unshaded part at higher Eddington ratio should
exhibit outflows, or the absorbing gas be far away from the nucleus
where the retaining gravitational mass is much larger.  They could for
instance be associated with dust lanes, as envisaged by
\protect\cite{2000A&A...355L..31M}. Such absorption cannot be too
large or the gas mass required would be prohibitive. We show, for
example, a limit at $N_{\rm H}=5\times 10^{21}\psqcm$ in Fig.~2.

Our prediction is therefore that absorbed objects should lie above the
approximate dividing line given by $N_{\rm H}>5\times 10^{23}\lambda
\psqcm$, provided that $N_{\rm H}>5\times 10^{21}\psqcm$.

\section{Comparison with data }

The simple model described above predicts that most highly-obscured
AGN will be observed to have an intrinsic column density placing them
in the shaded region of Fig. 2.

In order to test our model, we have examined the absorption and black
hole mass of several samples.  The first is a composite low redshift
sample consisting of the \protect\cite{2005ApJ...633L..77M} sample of
AGN detected by the Burst Alert Telescope (BAT) on the \emph{SWIFT}
satellite, and the \protect\cite{2007A&A...461.1209D} sample of
Seyfert nuclei observed by the \emph{BeppoSAX} satellite. The former
is an all-sky, hard-X-ray-flux limited sample in the energy range
14--195 keV where detection should be independent of column density,
provided the source is not too Compton thick. The latter study
presents an atlas of X-ray spectra for 39 Seyfert I and 42 Seyfert II
nuclei over the energy range 2--100 keV. All objects in the combined
sample have redshifts below 0.05. Black hole masses are estimated from
velocity dispersions using the $M-\sigma$ relation of
\protect\cite{2002ApJ...574..740T}.  Velocity dispersions were
obtained from the HyperLEDA online database.  We obtained 2-10keV
luminosities from the
Tartarus\footnote{http://astro.ic.ac.uk/Research/Tartarus/} database
of \emph{ASCA} observations, as it was not possible to extrapolate a
2-10keV luminosity from the SWIFT luminosities provided in
\protect\cite{2005ApJ...633L..77M} without a value for the photon
index $\Gamma$.  Finally, we convert to a bolometric luminosity using
the Eddington-ratio-dependent bolometric correction scheme of
\protect\cite{2007MNRAS.381.1235V}, using the black hole mass to
estimate the Eddington ratio $\lambda$, and assume that the emission
is isotropic.  We obtained velocity dispersions, and thus calculated
masses, for 23 objects in the composite sample.  The results for this
sample are shown in Fig. 3.

\begin{figure} 
\includegraphics[width=\columnwidth]{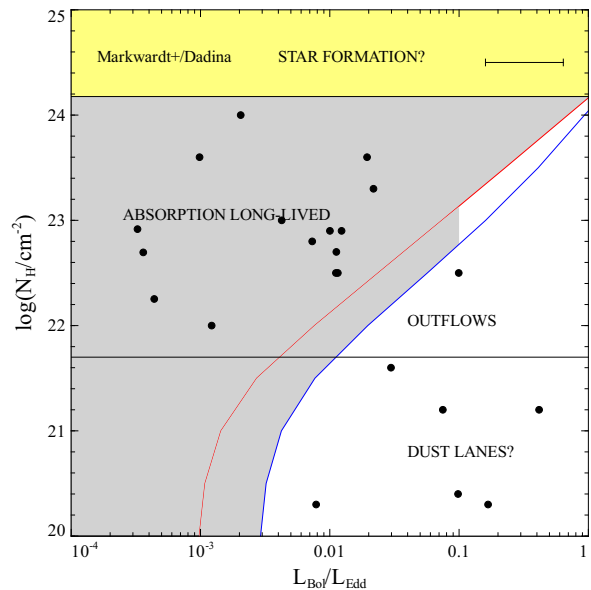}
\caption{Objects from the samples taken from the work of
  \protect\cite{2005ApJ...633L..77M} and
  \protect\cite{2007A&A...461.1209D} plotted on the $N_{\rm H} -
  \lambda=L_{\rm Bol}/L_{\rm Edd}$ plane. An estimate of the
  uncertainty in $\lambda=L_{\rm Bol}/L_{\rm Edd}$ is shown by the bar
  at top right. }
\end{figure}

In this Figure we see one object with a large column density and high
Eddington ratio in the unshaded region.  This is NGC 3783, and the
absorption is part of an outflowing warm absorber
\protect\citep{2001ApJ...554..216K}. The 3 objects with $22 > \log
N_{\rm H} > 21$ (just below the horizontal line) are, in order of
increasing $\lambda$, IC\,4329A, which has an outflow and is seen
almost edge on so absorption may be from a distant dust lane \citep{2006ApJ...646..783M}; NGC3516, which has
variable absorption and an outflow \citep{2007arXiv0710.0382M}; and 3C120, which has a soft excess in XMM spectra \citep{2004MNRAS.354..839B} so may have $N_{\rm
  H}$ overestimated in the value tabulated by Markwardt et al (2005)
used in Fig.~3.

We then used deeper samples.  First we use the Chandra Deep Field
South results of \protect\cite{2006A&A...451..457T}.  These authors
provide values for the column density $N_{\rm H}$ for each AGN
together with intrinsic X-ray luminosities.  We calculate black hole
mass estimates using K-band magnitudes from
\protect\cite{2004ApJS..155..271S} and the $M_{\rm BH}-L_{\rm K}$
relation of \protect\cite{2003ApJ...589L..21M}.  We impose a redshift
cut, requiring our objects to lie between redshifts $0.5<z<1.0$.  We
attempt to account for some evolution in the $M_{\rm BH}-L_{\rm K}$
relation between that epoch and our own by incrementing the K-band
magnitudes by unity before using the relation (i.e. accounting for
fainter bulge luminosities for the same central black hole mass in the
past).  This is broadly consistent with the evolution expected from 
\cite{2006ApJ...636L..21V}, assuming negligible evolution in the ratio
$M_{\rm{stellar}}/M_{\rm{BH}}$. This produces estimates for $M_{\rm BH}$ in line with other
measures.  Using bolometric corrections as for the local sample, we
estimate the Eddington ratios as before.  This sample yields 77
objects for which mass estimates could be obtained.  The results are
shown in Fig.~4 (top); objects with negligible column density are
marked nominally as having an upper limiting $N_{\rm H}$ of
$10^{21}\psqcm$ (downward arrows). The range of Eddington ratios that we find for the
CDFS sources is comparable to that found independently by
\protect\cite{2007A&A...474..755B}.
 
Finally we used the Lockman Hole results of
\protect\cite{2002A&A...393..425M} and
\protect\cite{2005A&A...444...79M}.  Mass estimates were calculated
using K-band magnitudes from \protect\cite{2002A&A...393..425M},
whereas X-ray luminosities and column densities were taken from
\protect\cite{2005A&A...444...79M} and bolometric corrections as
before.  The redshift range used was the same as for the CDFS sample.
There were 13 objects satisfying these criteria, and the results are
shown in Fig.~4 (bottom).

The results from the deeper samples are qualitatively similar to the
low redshift sample.  For all three samples we find, as predicted,
that most objects lie within the shaded region. Note that since
Chandra probes deeper than XMM, most of the CDFS objects are at a
lower value of $\lambda$ than those in the Lockman Hole.

We show relatively conservative estimates for the uncertaintiess
involved to provide some indication of the validity of our results.
In the local Seyfert sample (Fig.~3), we use published errors on
velocity dispersions and assume errors of 10 per cent on the Tartarus
X-ray fluxes and also on the values of $N_{H}$ provided.  For the
distant samples, we assume errors of 20 per cent on both K-band and
2--10 keV luminosities.  In the case of the K-band, we note that
Marconi and Hunt (2003) identify errors of $\pm 0.1$ on K-band
magnitudes, translating to $\sim 10$ per cent in luminosity; we
suggest that the 20 per cent value used here is therefore a reasonably
`safe' estimate of the error.  For the X-ray band, one may expect
variability to introduce significant uncertainties, but we suggest
that 20 per cent error is again relatively conservative for most
objects.  Such error estimates lead to uncertainties of around $\sim
40$ percent in the Eddington ratios; however, this is small enough to
clearly differentiate those lying within the shaded regions on Figs. 3
and 4 from those which are not.  The typical error in $\lambda$ under
these assumptions is shown in Figs. 3 and 4 by the bar in the top
right.

\begin{figure} 
\includegraphics[width=\columnwidth]{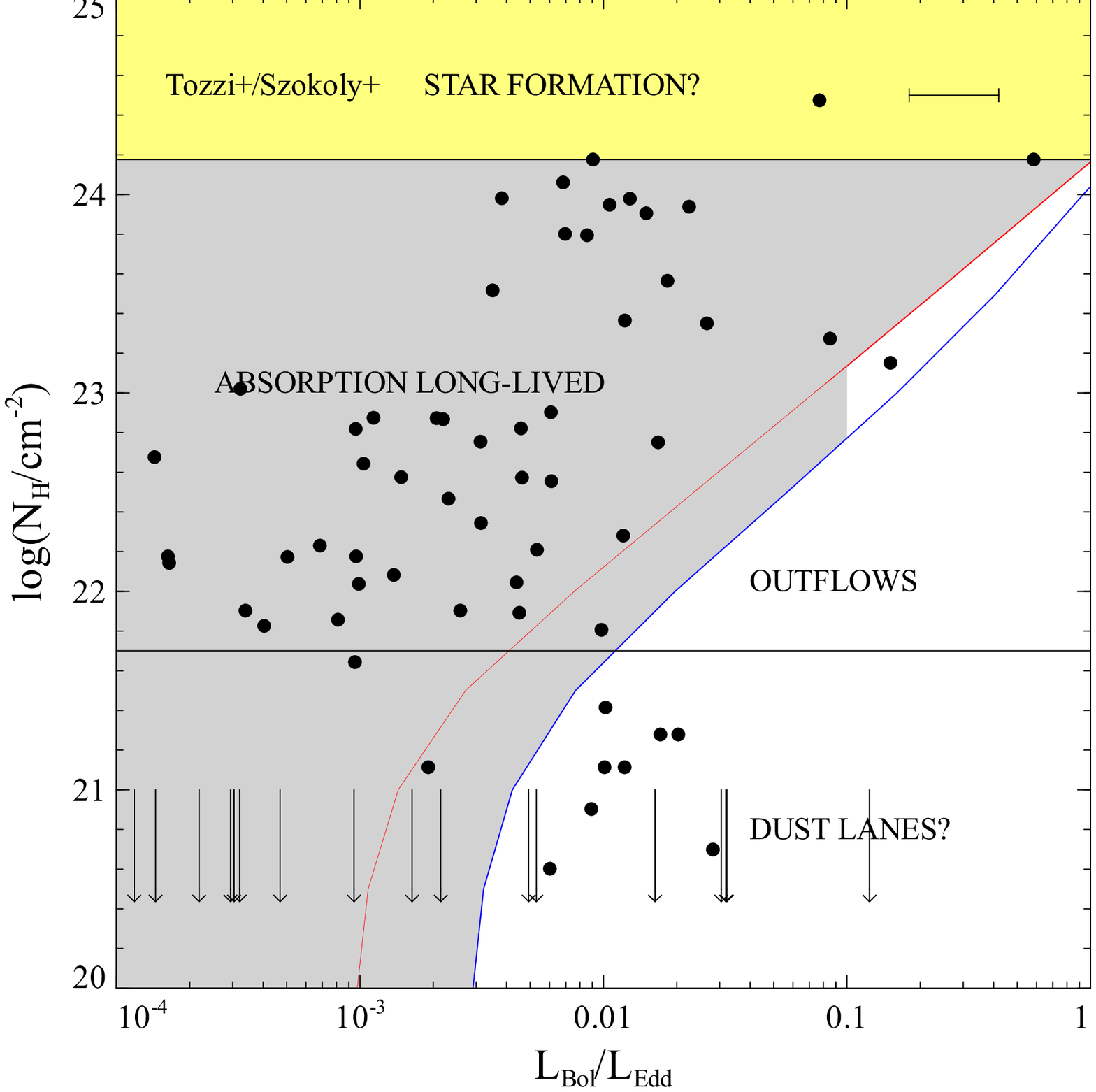} 
\includegraphics[width=\columnwidth]{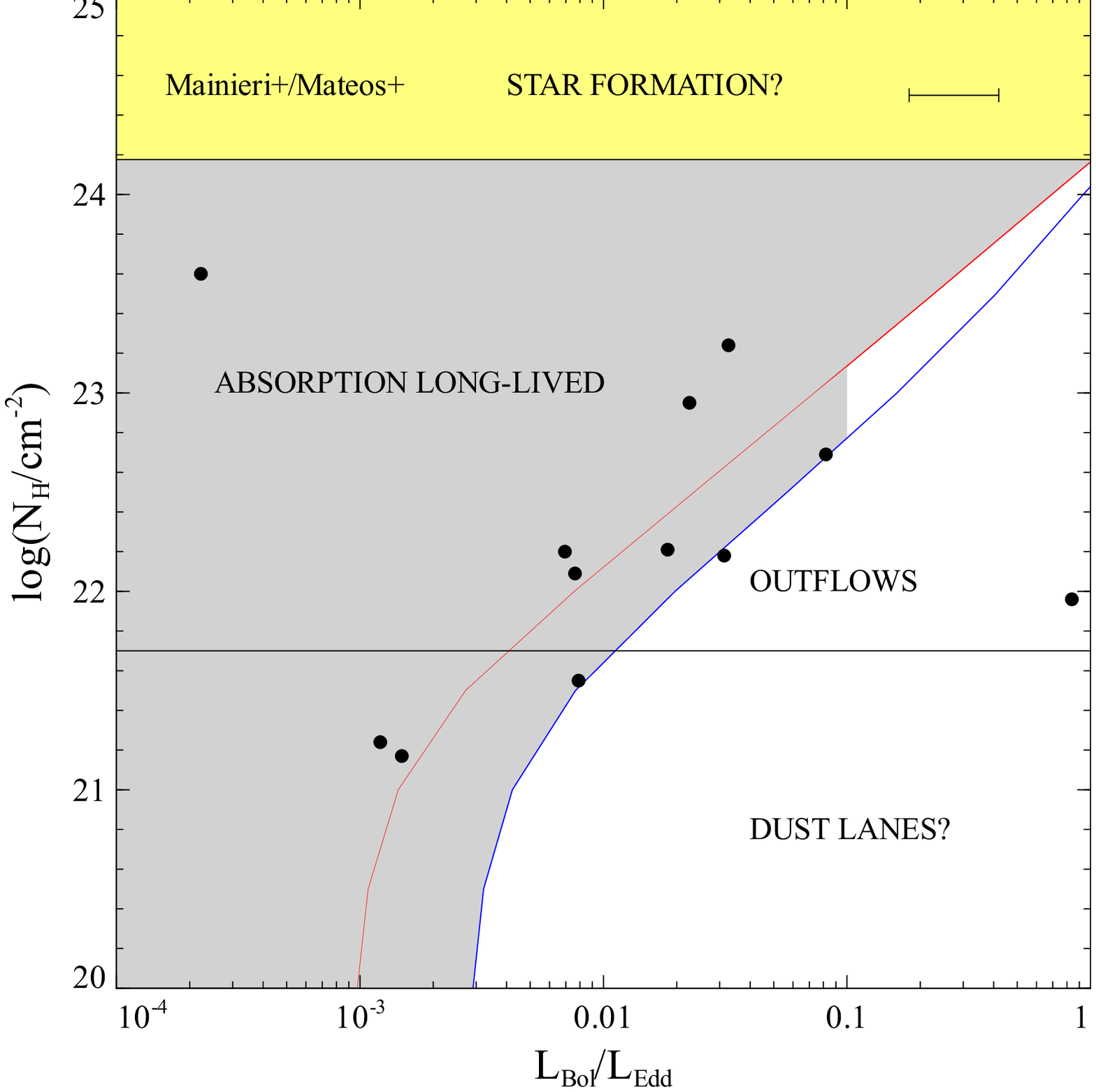}
\caption{The $N_{\rm H} - \lambda=L_{\rm Bol}/L_{\rm Edd}$ plane with
  objects from the CDFS (top) using data from
  \protect\cite{2006A&A...451..457T} and
  \protect\cite{2004ApJS..155..271S}), and the Lockman Hole (bottom)
  using data from \protect\cite{2002A&A...393..425M} and
  \protect\cite{2005A&A...444...79M}. There is one point missing from
  each plot with $\lambda<10^{-4}$. }
\end{figure}

\section{Discussion}

We see from the plots that most absorbed AGN with column densities
$N_{\rm H}>10^{22}\pcmsq$ have an Eddington ratio $\lambda <0.1$, as
expected if radiation pressure on dusty gas is important. The observed
objects are sub-Eddington in terms of radiation pressure on dusty gas
and so the absorbing gas may be long-lived.  The line defining the
effective Eddington limit in the plots is determined for the mass of
the black hole only. It shifts to the right (higher values of
$\lambda$) if larger masses associated with the stellar bulge are also
involved.  Since the lines drawn seem to apply to the samples, most of
the absorbing gas must be at small radii (much less than 100~pc). The
agreement between the expected location of absorbed AGN and the
observed locations provides support for the idea that radiation
pressure acts to blow gas away from galaxy bulges. If the radiation is
isotropic then it can stem the growth of both inner bulge and black
hole.
 
The inner gas directly exposed to the radiation pressure may be
unstable to clumping (\citealt{1979ApJ...233..479B}; \citealt{2007MNRAS.380.1172H}),
 the details of which will not be pursued here. Most of the
long-lived gas at column densities above $10^{22}\psqcm$ is shielded
by the inner gas and so need not be clumped.

After blowing away the gas, AGN may decline in luminosity
so creating unabsorbed sources at low Eddington ratios. There is also
some uncertainty in the boundary between the various regions of the
diagrams due to factors such as source variability and clumpiness in the
absorption. The fact that most sources avoid the region of our
diagrams above $N_{\rm H}\sim 5\times 10^{21}\psqcm$ and to the right
of the radiation pressure line demonstrates that variabilitity
is not very important. 

Lower levels of absorption below $10^{22}\pcmsq$ can be long-lived at
large radii in a galaxy, since the relevant gravitating mass there is
due to the black hole and the bulge. The radiation limit in our
Figures shifts to the right by a factor of $M_{\rm bulge}/M_{\rm BH}$
with all gas above the line bound to the bulge.  The maximum boost $A$
that can be obtained from radiation pressure on dusty gas is
$\sigma_{\rm d}/\sigma_{\rm T}\sim 500$ which means that the black
hole is above the effective Eddington limit for the whole bulge if
$M_{\rm BH}/M_{\rm bulge}>1/500$.

Consequently we envisage a scenario where a black hole smothered in
gas could grow in a bulge in stages. It pushes the gas out to a
distance in the bulge where the mass within that radius is 500 times
the black hole mass. (The boost factor increases as the column density
decreases, so once gas starts to move outward it continues to do so,
see Fig.~1 and \protect\citealt{2006MNRAS.373L..16F}) After the
accretion disk empties, the AGN switches off. If the bulge mass within
the radius to which the gas was pushed exceeds 500 times the mass of
the black hole, then the gas falls back in and the cycle
repeats. Through accretion, the black hole mass and thus luminosity
increases each cycle until is unable to retain the gas and it is
pushed right out of the bulge. At this
point
\begin{equation}
  M_{\rm BH}/M_{\rm bulge}\sim \sigma_{\rm T}/\sigma_{\rm d}\sim 1/500,
\end{equation}
similar to the value found by Marconi \& Hunt (2004) from correlating
the observed properties of galaxies.

Star formation from the gas in the galaxy during these cycles
presumably leads to
the bulge mass -- velocity dispersion (Faber--Jackson) relation
required such that equations (1), which acts locally, and (2), which
acts globally, agree. 

For much of the `cycling scenario' envisaged above, the only
acceptable range for bright unabsorbed objects would be at high
Eddington ratios.  From a large sample of AGN detected in their AGN
and Galaxy Evolution Survey (AGES), \cite{2006ApJ...648..128K} find
$\lambda\sim 0.2$. An important result from their survey is
that they should have been sensitive to unabsorbed AGN with lower
Eddington ratios, but found none. This could in part be a selection
effect due to absorption since the Chandra X-ray observations used are
short, about 5~ks, which means that they are most sensitive to bright
unabsorbed objects.  The discussion of emission line strength for high and
low Eddington ratio AGN from \cite{2007MNRAS.381.1235V} could also be of 
particular relevance here, again implying that higher Eddington ratio AGN
 would be systematically favoured. The detected objects are in the unshaded part of
our diagram, so have high $\lambda$.

\section{Summary}

Absorbed AGN are most commonly found at low Eddington ratios such that
they are sub-Eddington for dusty gas. This agrees with the hypothesis
that radiation pressure acting on dust is important in removing gas
from galaxy bulges. In turn this leads to $M_{\rm BH} - \sigma$ and
$M_{\rm BH} - M_{\rm bulge}$ relations similar to those observed.

Studies seeking to examine the evolution of absorbed AGN will need to
include the dependence on Eddington ratio. 

\section{Acknowledgments} We acknowledge Gary Ferland for the use of
his code {\sc cloudy} and for discussions. We thank Silvia Mateos and
Vincenzo Mainieri for helpful suggestions and information on the data
presented in their papers. ACF thanks The Royal Society for support,
RV acknowledges support from the UK Science and Technology Funding
Council (STFC) and PG acknowledges a Fellowship from the Japan Society
for the Promotion of Science (JSPS).  This research has made use of
the Tartarus (Version 3.2) database, created by Paul O'Neill and
Kirpal Nandra at Imperial College London, and Jane Turner at
NASA/GSFC.  Tartarus is supported by funding from STFC (PPARC), and
NASA grants NAG5-7385 and NAG5-7067.

\bibliographystyle{mnras} 
\bibliography{eddabs}

\end{document}